\documentclass[aps,prb,twocolumn,showpacs,showkeys,preprintnumbers,groupedaddress,amsmath,amssymb,showpacs]{revtex4-1}

\usepackage{graphicx}
\usepackage{dcolumn}
\usepackage{bm}
\usepackage{hyperref}
\usepackage{color}

\begin{document}

\preprint{\textit{Published in} Nanotechnology \textbf{24}, 245203 (2013)}


\title{Quantum-interference transport through surface layers of indium-doped ZnO nanowires}

\author{Shao-Pin Chiu,$^1$, Jia Grace Lu,$^2$, Juhn-Jong Lin$^{1,3,\ast}$}
\affiliation{$^1$Institute of Physics, National Chiao Tung University, Hsinchu 30010, Taiwan \\
$^2$Department of Physics \& Astronomy, University of Southern California, Los Angeles, California
90089-0484, USA \\ $^3$Department of Electrophysics, National Chiao Tung University, Hsinchu 30010,
Taiwan}


\begin{abstract}

We have fabricated indium-doped ZnO (IZO) nanowires (NWs) and carried out four-probe
electrical-transport measurements on two individual NWs with geometric diameters of $\approx$ 70 and
$\approx$ 90 nm in a wide temperature $T$ interval of 1--70 K. The NWs reveal overall charge
conduction behavior characteristic of disordered metals. In addition to the $T$ dependence of
resistance $R$, we have measured the magnetoresistances (MR) in magnetic fields applied either
perpendicular or parallel to the NW axis. Our $R(T)$ and MR data in different $T$ intervals are
consistent with the theoretical predictions of the one- (1D), two- (2D) or three-dimensional (3D)
weak-localization (WL) and the electron-electron interaction (EEI) effects. In particular, a few
dimensionality crossovers in the two effects are observed. These crossover phenomena are consistent
with the model of a ``core-shell-like structure" in individual IZO NWs, where an outer shell of a
thickness $t$ ($\simeq$ 15--17 nm) is responsible for the quantum-interference transport. In the WL
effect, as the electron dephasing length $L_\varphi$ gradually decreases with increasing $T$ from the
lowest measurement temperatures, a 1D-to-2D dimensionality crossover takes place around a
characteristic temperature where $L_\varphi$ approximately equals $d$, an effective NW diameter which
is slightly smaller than the geometric diameter. As $T$ further increases, a 2D-to-3D dimensionality
crossover occurs around another characteristic temperature where $L_\varphi$ approximately equals $t$
($< d$). In the EEI effect, a 2D-to-3D dimensionality crossover takes place when the thermal diffusion
length $L_T$ progressively decreases with increasing $T$ and approaches $t$. However, a crossover to
the 1D EEI effect is not seen because $L_T < d$ even at $T$ = 1 K in our IZO NWs. Furthermore, we
explain the various inelastic electron scattering processes which govern $L_\varphi$. This work
demonstrates the complex and rich nature of the charge conduction properties of group-III metal doped
ZnO NWs. This work also strongly indicates that the surface-related conduction processes are essential
to doped semiconductor nanostructures.

\end{abstract}

\pacs{73.63.-b, 73.20.Fz, 72.20.-i, 72.80.Ey}
\maketitle

\section{Introduction}

In recent years, nanometer-scale structures have opened up numerous new horizons in both fundamental
and applied research \cite{Nazarov}. Among the various forms of nanoscale structures, nanowires (NWs)
provide the unique advantages for facilitating four-probe electrical-transport measurements over a
wide range of temperature $T$ and in externally applied magnetic fields $B$. In-depth investigations
of the rich phenomena and the underlying physics of intrinsic charge and spin conduction processes in
single NWs are thus feasible. Indeed, to date, significant advances have been made in studies of
metallic \cite{Chiquito07,LinYH-nano08,Chiu-nano09ITO,Hsu-prb10,YangPY-prb12}, magnetic
\cite{Tatara08}, semiconducting \cite{Rueb-prb07,Chiu-nano09ZnO, Tsai-nano10, Petersen09,Liang10,
Hau10,Xu10,Hernandez10,Zeng-nanolett12}, and superconducting \cite{Arutyunov08} NWs.

Zinc oxide (ZnO) NWs are probably the most extensively studied materials among all kinds of
semiconductor NWs, due to their intricate physical properties as well as their widespread potential
applications in nanoelectronic and spintronic devices \cite{Ozgur05,Wang08}. In addition to the
initial investigations of natively (unintentionally) doped samples, artificially doped ZnO NWs have
attracted much attention. For instance, the n-type doping of Ga and In, among other metal atoms, into
ZnO NWs has recently been explored \cite{Liu10,Ahn07}. The electrical-transport properties of
individual In-doped ZnO (hereafter, referred to as IZO) NW \cite{JGL09} and NW transistors \cite{Xu10}
have also been reported. Despite the intense experimental studies in the past years, the electrical
conduction mechanisms in the parent ZnO NWs have only been explained recently. Chiu {\it et al.}
\cite{Chiu-nano09ZnO} and Tsai {\it et al.} \cite{Tsai-nano10} have demonstrated that most
artificially synthesized ZnO NWs, being essentially independent of the growth method, are moderately
highly doped and weakly self-compensated, resulting in a splitting of the impurity band. As a result,
the overall charge transport behavior is due to the ``split-impurity-band conduction" processes.
Furthermore, Chiu {\it et al.} \cite{Chiu-nano09ZnO} and Tsai {\it et al.} \cite{Tsai-nano10} have
shown that many natively doped ZnO NWs, which inherently possess high carrier (electron)
concentrations, often lie on the insulating side of, but very close to, the metal-insulator (M-I)
transition. Therefore, it is expected that incorporation of a few atomic percent of, e.g., the
group-III indium atoms into ZnO NWs may promote the NWs to fall on the metallic side of the M-I
transition. (The group-III atoms, Al, Ga, and In, are shallow donors in ZnO \cite{Look08}.) Thus, the
IZO NWs should reveal electrical conduction properties characteristic to those of disordered
conductors. In particular, the quantum-interference weak-localization (WL) and the electron-electron
interaction (EEI) effects should be manifest at low temperatures
\cite{Bergmann84,Bergmann10,Altshuler85}.

Apart from the interesting low-dimensional electrical-transport properties that could be expected for
NW structures, ZnO NWs inherit some complexities, as compared with other NW materials. In particular,
the question of whether the surfaces of a ZnO NW are more conducting or less conducting than the bulk
has been investigated by several groups. It is now accepted that surface electron accumulation layers,
band bending effects, compositional nonstoichiometries, and ambient conditions, etc., can all markedly
affect the electrical properties of a surface \cite{Schlenker08,Allen10}. Recent measurements of
resistance $R$ and magnetoresistance (MR) by Chiu {\it et al.} \cite{Chiu-nano09ZnO}, Hu {\it et al.}
\cite{Hu09}, and Tsai {\it et al.} \cite{Tsai-nano10} have revealed that surface conduction is
particularly important in those ZnO NWs lying close to the M-I transition. In this context, it would
be very interesting to investigate if surface conduction could also be pronounced in IZO NWs which are
even more metallic than the natively doped ZnO NWs. Indeed, in this work, we have measured and
analyzed the $T$ dependence of $R$ as well as the $B$ dependence of both perpendicular MR and parallel
MR in two IZO NWs over a wide $T$ interval of 1--70 K. We found a few dimensionality crossovers in the
WL and the EEI effects as $T$ gradually increases from 1 to 70 K. These results strongly point to
dominating roles of the {\em surface-related conduction} processes, prompting us to propose a
``core-shell-like structure" in individual IZO NWs. This present work demonstrates the complex and
rich nature of charge transport processes in ZnO-based NWs. This work also illustrates that
quantum-interference transport studies can provide a useful probe for the electron scattering
processes in these nanoscale materials. Furthermore, regarding to the materials and technological
aspects, we would like to stress that the nature of impurity doping semiconductor nanostructures is
\textit{intrinsically distinct} from doping the bulks. The recent theoretical calculations of Dalpian
and Chelikowsky \cite{Dalpian06} have shown that {\em dopants would be energetically in favor of
segregating to the surfaces rather than distributing uniformly across the radial direction}. Their
theory provides a strong microscopic support for our experimental observation of a core-shell-like
structure in the IZO NWs.

\begin{figure}[htp]
\begin{center}
\includegraphics[scale=0.20]{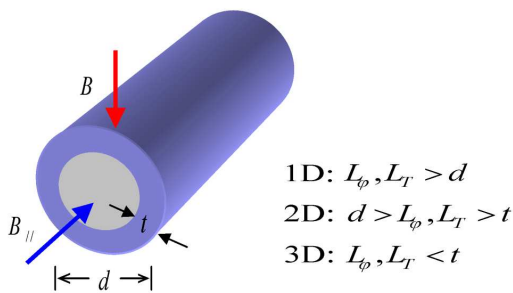}
\caption{A schematic for our proposed core-shell-like structure in IZO NWs. The blue outer
shell of a thickness $t$ is responsible for the WL and the EEI transport. $d$ is an effective
diameter which differs slightly from the geometric diameter. The experimental conditions for
the 1D, 2D, and 3D WL and EEI effects are summarized on the right. \label{fig1}}
\end{center}
\end{figure}

For the convenience of discussion, we first present our model of the core-shell-like structure and
summarize the main results of this work. Figure \ref{fig1} shows a schematic of the core-shell-like
structure with a surface conduction layer of a thickness $t$ ($\simeq$ 15-17 nm in our IZO NWs). The
sizes of the effective NW diameter $d$ and the thickness $t$, relative to the electron dephasing
length $L_\varphi$ and the thermal diffusion length $L_T$, determine the observed one-dimensional
(1D), two-dimensional (2D), or three-dimensional (3D) WL and EEI effects, as summarized in
figure~\ref{fig1}. The dephasing length $L_\varphi = \sqrt{D \tau_\varphi}$ is the characteristic
length scale in the single-particle WL effect and the thermal length $L_T = \sqrt{D \hbar /k_BT}$ is
the characteristic length scale in the many-body EEI effect, where $D$ is the diffusion constant,
$\tau_\varphi$ is the electron dephasing time, $2 \pi \hbar$ is the Planck constant, and $k_B$ is the
Boltzmann constant. Table \ref{t1} summarizes the various $T$ intervals over which different
dimensionalities in the WL and the EEI effects are observed in our IZO NWs.

\begin{table*}
\caption{\label{t1} Summary of approximate temperature intervals over which different dimensionalities
in the WL and the EEI effects are observed in IZO NWs.}

\begin{ruledtabular}
\begin{tabular}{lcccccccc}
Nanowire & & WL effect & & & & & EEI effect & \\  \cline{2-4}  \cline{7-9}
 & 1D & 2D & 3D & & & 1D & 2D & 3D \\
\hline
IZO1 & $\sim$ 1--7 K & $\sim$ 7--50 K & $\gtrsim$ 50 K & & & --- & $\sim$ 3--11 K & $\gtrsim$ 14 K \\
IZO2 & --- & $\sim$ 1--40 K & $\gtrsim$ 40 K & & & --- & --- & $\gtrsim$ 3 K \\
\end{tabular}
\end{ruledtabular}
\end{table*}

This paper is organized as follows. In section 2, we discuss our experimental method for the NW
synthesis and characterizations as well as the low-$T$ four-probe resistance and MR measurements. In
sections 3 to 5, we present our experimental results and discussions in great detail. We carry out
analyses of the perpendicular MR data and the parallel MR data (section 3) as well as the $T$
dependence of $R$ (section 4) to illustrate how a conducting outer-layer must exist in single IZO NWs.
As a consequence, a few 1D-to-2D and 2D-to-3D dimensionality crossovers in the WL and the EEI effects
are observed. In section 5, we identify the underlying electron dephasing processes. Our conclusion is
given in section 6.

\section{Experimental method}

\begin{figure}[htp]
\begin{center}
\includegraphics[scale=0.23]{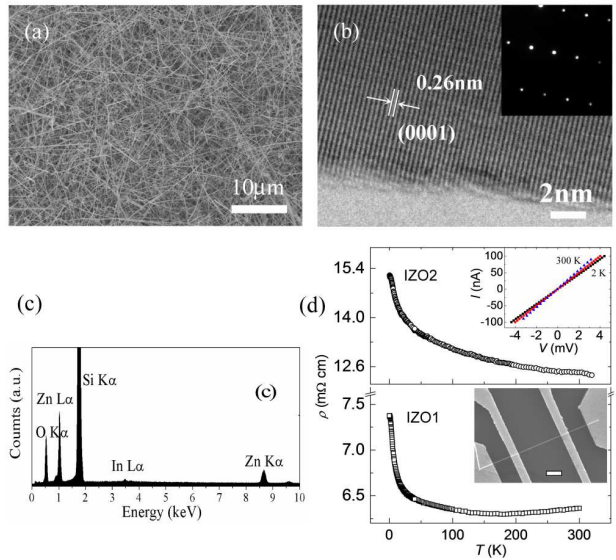}
\caption{(a) An SEM image of as-grown IZO NWs on a tin-coated Si substrate. (b) An HRTEM image of a single IZO NW.
The inset shows the corresponding selected-area electron diffraction pattern. (c) EDX spectrum of as-grown IZO NWs.
(d) Resistivity $\rho$ as a function of temperature for the IZO1 and IZO2 NWs, as indicated. The inset in the upper
panel shows the $I$-$V$ curves for the IZO2 NW at 2, 40 and 300 K. Note that these $I$-$V$ curves are linear.
They nearly overlap because the resistivity of this given NW depends weakly on $T$ (main panel). The inset in the
lower panel shows an SEM image of the IZO1 NW device. The scale bar is 1 $\mu$m. \label{fig2}}
\end{center}
\end{figure}

Our IZO NWs were synthesized by the laser-assisted chemical vapor deposition (CVD) method, as
described previously \cite{JGL09}. The scanning electron microscopy (SEM) image in
figure~\ref{fig2}(a) shows that IZO NWs with diameters ranging roughly from 40 to 100 nm were formed
on a tin-coated Si substrate. The high-resolution transmission electron microscopy (HRTEM) image and
selected-area electron diffraction pattern in figure~\ref{fig2}(b) indicate that In atoms were
effectively incorporated into the wurtzite crystal lattice of ZnO and the single crystalline structure
with the (0001) growth direction was maintained. Energy-dispersive x-ray (EDX) spectrum displayed in
figure~\ref{fig2}(c) indicates an In to Zn atomic ratio of approximately 3 at.\%, suggesting
successful incorporation of In atoms into the ZnO crystal lattice. The NW growth and doping method as
well as the structural and composition analyses were previously discussed in reference
\onlinecite{JGL09}.

We have fabricated several single IZO NW devices with the four-probe configuration by utilizing the
electron-beam lithography technique. The devices reported in this paper were taken from the same batch
of NWs. Submicron Cr/Au (10/100 nm) electrodes were made via thermal evaporation deposition. The inset
of the lower panel in figure~\ref{fig2}(d) shows an SEM image of the IZO1 NW device. The contact
resistance between an electrode and the NW is typically a few k$\Omega$ at 300 K and $\sim$ 20
k$\Omega$ at 4 K. Extensive measurements of the resistances and magnetoresistances have been carried
out on two devices. The experimental setup and measurement procedures were similar to those employed
in our previous studies of single natively doped ZnO NWs \cite{Chiu-nano09ZnO,Tsai-nano10} and indium
tin oxide (ITO) NWs \cite{Hsu-prb10}. The $R(T)$ and MR curves were measured by utilizing a Keithley
K-220 or K-6430 as a current source and a high-impedance (T$\Omega$) Keithley K-2635A or K-6430 as a
voltameter. The resistances reported in this work were all measured by scanning the current-voltage
($I$-$V$) curves at various fixed $T$ values between 300 and 1 K. The resistance at a given $T$ value
was then determined from the region around the zero bias voltage, where the $I$-$V$ curve was
definitely linear, see the inset in the upper panel of figure~\ref{fig2}(d). In fact, since our IZO
NWs had relatively low resistivities ($\sim$ 10 m$\Omega$ cm) which depended very weakly on $T$ in the
wide temperature range 1--300 K, the NWs were ``metallic-like" and the electron-beam lithographic
contacts were already ohmic without any heat treatment. Electron overheating at our lowest measurement
temperatures was carefully monitored and largely avoided except that there might be slight heating for
those data points taken at $T$ = 1 K. This possible slight electron heating at 1 K will not affect any
of our discussions and conclusion, except the extracted value of $\tau_\varphi^{-1}$(1\,K) might be
slightly overestimated (figure~\ref{fig7}). Notice that, since we had employed the four-probe
configuration, the measured resistances (resistivities) were thus the intrinsic resistances
(resistivities) of the individual NWs. The relevant parameters of the two individual IZO NW devices
studied in this work are listed in table \ref{t2}.

\begin{table*}
\caption{\label{t2} Relevant parameters of two single IZO NW devices. $dia$ is the geometric diameter
measured by SEM, $L$ is the length between the two voltage leads in a four-probe configuration, $\rho$
is the resistivity, $n$ is the carrier concentration, $E_F$ ($k_F$) is the Fermi energy (wave number),
$\ell$ ($\tau_e$) is the elastic mean-free path (time), and $D$ is the diffusion constant. $\ell$,
$\tau_e$, $D$, and $k_F \ell$ values are for 10 K.}

\begin{ruledtabular}
\begin{tabular}{lccccccccccc}
Nanowire & $dia$ & $L$ & $R$(300\,K) & $\rho$(300\,K) & $\rho$(10\,K) & $n$ & $E_F$ & $\ell$ & $\tau_e$ & $D$ & $k_F \ell$ \\
 & (nm) & ($\mu$m) & (k$\Omega$) & (m$\Omega$ cm) & (m$\Omega$ cm)  & (cm$^{-3}$) & (meV) & (nm) & (fs) & (cm$^2$/s) & \\
\hline
IZO1 & 68 & 3.8 & 66.6 & 6.4 & 6.7 & 1.7$\times$$10^{19}$ & 100 & 2.8 & 7.3 & 3.6 & 2.2 \\
IZO2 & 92 & 1.9 & 35.5 & 12 & 15 & 6.8$\times$$10^{18}$ & 55 & 2.4 & 8.4 & 2.2 & 1.4 \\
\end{tabular}
\end{ruledtabular}
\end{table*}

\section{Results and discussion: magnetoresistance in the weak-localization effect}

This section is divided into three subsections. In subsection 3.1, we present our estimates of the
electronic parameters in our IZO NWs. In subsection 3.2, we present our perpendicular MR data and the
observed dimensionality crossovers in the WL effect. In subsection 3.3, we analyze our parallel MR
data to further determine and confirm the thickness of the outer conduction shell in single IZO NWs.

\subsection{Estimate of nanowire electronic parameters}

In order to facilitate quantitative comparison of our $R(T)$ and MR data with the WL and the EEI
theoretical predictions, we first discuss the estimates of the relevant electronic parameters of our
IZO NWs. The carrier concentration $n$ of a doped semiconductor NW can not be readily measured, e.g.,
by using the conventional Hall effect, due to the small transverse dimensions of a single NW.
Fortunately, insofar as single-crystalline ZnO materials (films and bulks) are concerned, an empirical
relation between the room-temperature resistivity $\rho$(300 K) and $n$ has been comprehensively
compiled and reliably established by Ellmer in the figure 4 of reference \onlinecite{Ellmer01}.
Furthermore, Chiu {\it et al.} \cite{Chiu-nano09ZnO} and Tsai {\it et al.} \cite{Tsai-nano10} have
recently shown that this Ellmer $\rho$--$n$ empirical relation can well be extended to the case of
individual single-crystalline ZnO NWs. Therefore, in this study we have applied this empirical
relation to evaluate the $n$ values in our IZO NWs.

It is worth noting that the $n$ values we evaluated (table \ref{t2}) are in reasonable consistency
with that extracted from direct measurements by using the back-gate method \cite{JGL09}. For example,
in an IZO NW with $\rho$(4 K) = 2.7 m$\Omega$ cm, the back-gate method reported an estimate of $n
\approx$ 1.2$\times$$10^{20}$ cm$^{-3}$. Alternatively, according to the Ellmer $\rho$--$n$ empirical
relation \cite{Ellmer01}, such a resistivity would infer a value of $n \approx$ 8$\times$$10^{19}$
cm$^{-3}$. That is, the $n$ values estimated according to the two independent methods agree to within
a factor of $\approx$ 1.5. Since the critical carrier concentration for the M-I transition in
single-crystalline ZnO occurs at  $n_c \approx$ 5$\times$$10^{18}$ cm$^{-3}$
\cite{Chiu-nano09ZnO,Tsai-nano10,Hutson57}, our doped IZO NWs obviously lie on the metallic side of,
but close to, the M-I transition. That our NWs lie close to the M-I transition boundary is directly
evident in the fact that our measured $T$ dependence of $\rho$ is weak, namely, $\rho$(1 K)/$\rho$(300
K) $\lesssim$ 1.2 in both NWs (figure~\ref{fig2}(d)). In the IZO1 NW, $\rho$ decreases with reducing
$T$ between 180 and 300 K, before it increases with further decrease in $T$. The metallic-like
behavior results from the incorporation of a few atomic percent of In atoms (donors) into the ZnO NW
crystal lattice as well as from the In doping induced oxygen vacancies \cite{Liu10}. The notable
resistivity rise below $\sim$ 40 K originates from the WL and the EEI effects. Numerically, the
$\rho$(300 K) values of the IZO1 and IZO2 NWs are approximately one order of magnitude lower than
those of the natively doped ZnO NWs that we had previously studied \cite{Chiu-nano09ZnO,Tsai-nano10}.
Moreover, these $\rho$(300\,K) values are $\sim$ 3 orders of magnitude lower than that in a two-probe
individual IZO NW transistor recently fabricated by Xu {\it et al.} \cite{Xu10}.

In estimating the other charge carrier parameters which are listed in table \ref{t2}, we have assumed
a free-electron model and taken an effective electron mass of $m^\ast = 0.24\, m$ in the conduction
band \cite{Baer67}, where $m$ is the free-electron mass. We obtain the electron mobility $\mu$(300\,K)
$\approx$ 55 cm$^2$/V\,s in the IZO1 NW and 75 cm$^2$/V\,s in the IZO2 NW. These values are slightly
lower than that ($\sim$ 100 cm$^2$/V\,s) found in natively doped ZnO NWs \cite{Chiu-nano09ZnO}. Note
that our evaluated Fermi energies $E_F$ are 100 and 55 meV in the IZO1 and IZO2 NWs, respectively.
These values are larger than the thermal energy $k_BT$ at 300 K, and also larger than the major
shallow donor level ($\sim$ 30 meV below the conduction band minimum
\cite{Chiu-nano09ZnO,Tsai-nano10,LienCC-jap11}) in the parent ZnO. Therefore, degenerate Fermi-liquid
and metallic behavior is to be expected in our NWs. As a consequence, all of our electronic parameters
depend only weakly on $T$, as compared with those in typical semiconductor NWs that exhibit, e.g.,
hopping conduction processes \cite{x4}.

\subsection{Perpendicular magnetoresistance in the weak-localization effect}

In disordered conductors and at low temperatures, the WL and the EEI effects cause pronounced
quantum-interference transport phenomena which depend sensitively on $T$ and $B$. The WL effect
results from the constructive interference between a pair of time-reversal partial electron waves
which traverse a closed trajectory in a random potential. The time-reversal symmetry will be readily
broken in the presence of a small $B$ field \cite{Bergmann84,Bergmann10}. The low-field MR can provide
quantitative information on the various electron dephasing mechanisms, such as the inelastic
electron-electron scattering, electron-phonon scattering, spin-orbit scattering, and magnetic
spin-spin scattering \cite{Lin-jpcm02}. The WL effects in different dimensionalities assume different
functional forms of MR \cite{Bergmann84,Bergmann10,Altshuler85}. Therefore, it is of crucial
importance to apply the appropriate MR expressions to describe the WL effect in those samples (e.g.,
NWs and thin films) whose transverse dimensions are comparable to the dephasing length $L_\varphi$. In
such cases, a dimensionality crossover of the WL effect is deemed to occur if the measurement $T$ is
varied sufficiently widely \cite{Mani-prb93}. Moreover, an external $B$ applied in the perpendicular
or the parallel orientation relative to the current flow can result in distinct MR, because the WL
effect is an orbital phenomenon in nature. Similarly, dimensionality crossovers could occur in the EEI
effect (section 4).

\begin{figure}[htp]
\begin{center}
\includegraphics[scale=0.23]{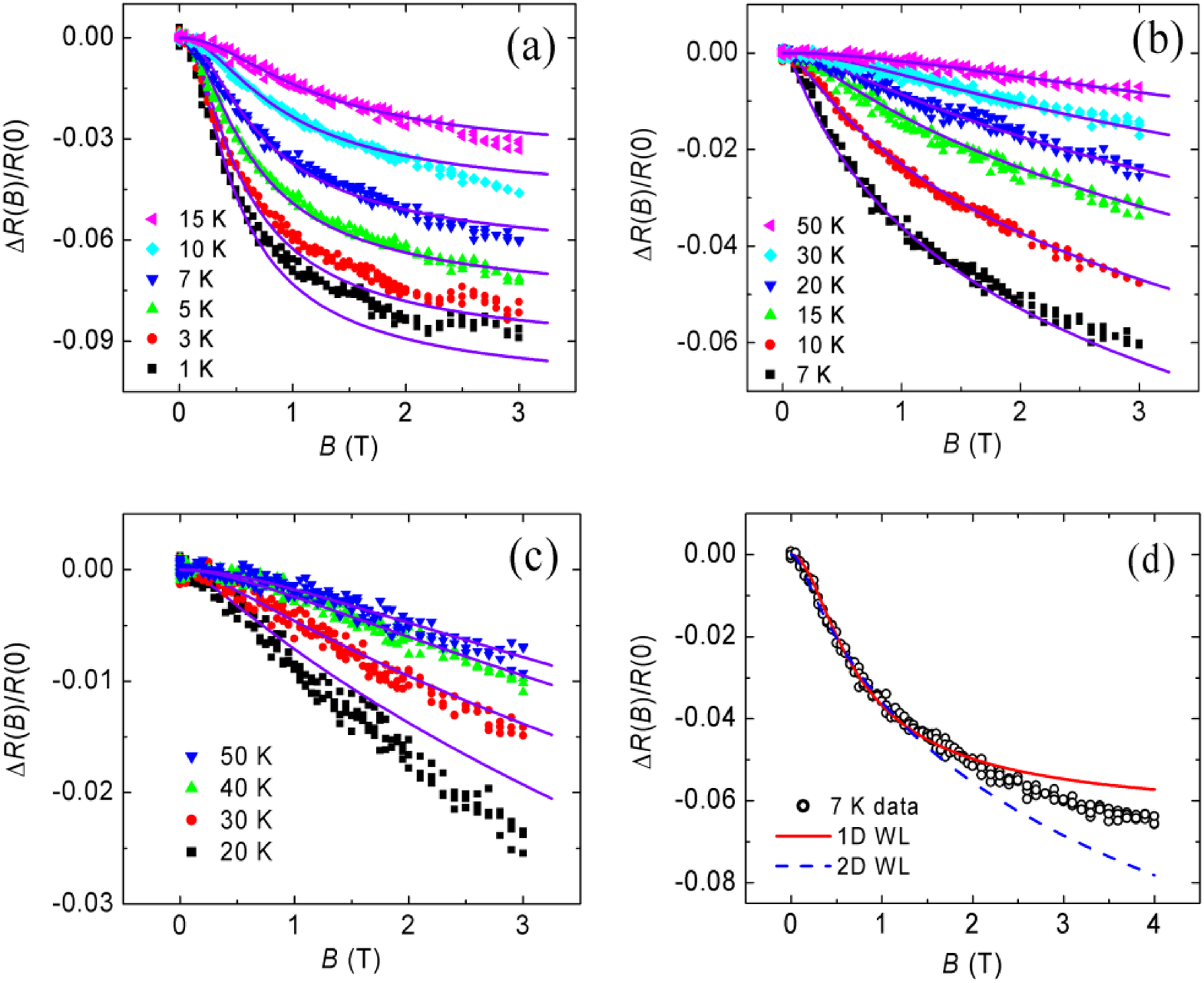}
\caption{Normalized MR, $\triangle R(B)/R(0) = (R(B) - R(0))/R(0)$, as a function of perpendicular
magnetic field of the IZO1 NW in different $T$ regions, as indicated. The symbols are the
experimental data and the solid curves are the theoretical predictions of the (a) 1D (equation~(\ref{1D})),
(b) 2D (equation~(\ref{2D})), and (c) 3D (equation~(\ref{3D})) WL effects. (d) $\triangle R(B)/R(0)$
at 7 K with the theoretical predictions of the 1D and 2D WL effects, as indicated.
\label{fig3}}
\end{center}
\end{figure}

Figures \ref{fig3}(a)--\ref{fig3}(d) show the normalized MR, $\triangle R(B)/R(0) = (R(B) -
R(0))/R(0)$, of the IZO1 NW as a function of perpendicular magnetic field in several $T$ regions, as
indicated. (The $B$ fields were applied perpendicular to the NW axis.) The symbols are the
experimental data and the solid curves are the WL theoretical predictions for 1D
(figure~\ref{fig3}(a)), 2D (figure~\ref{fig3}(b)), and 3D (figure~\ref{fig3}(c)), respectively. Figure
\ref{fig3}(d) shows a plot of the measured perpendicular MR at 7 K together with both the 1D and the
2D WL theoretical predictions, as indicated. We notice that in all figures
\ref{fig3}(a)--\ref{fig3}(d), the MR data are negative, suggesting that the spin-orbit (s-o)
scattering rate is relatively weak compared with the inelastic electron scattering rate at all $T$
down to 1 K. In other words, the s-o scattering length $L_{so} = \sqrt{D \tau_{so}}$ is always longer
than $L_\varphi$, where $\tau_{so}$ is the s-o scattering time. This observation of a very weak s-o
scattering is consistent with the conclusion recently drawn from the WL studies of metallic-like ZnO
NWs \cite{Chiu-nano09ZnO} and ZnO nanoplates \cite{Likovich09,Andrearczyk05}. Microscopically, the
comparatively weak s-o coupling in ZnO is thought to originate from the small energy splitting at the
top of the valence band \cite{Harmon09}. A doping of $\approx$ 3 at.\% of moderately heavy In atoms in
this work does not induce any appreciable enhancement of the s-o coupling.

The MR due to the 1D WL effect in the presence of an external $B$ applied either perpendicular or
parallel to the NW axis is given by \cite{Altshuler81,Birge03}, in terms of the normalized resistance
$\triangle R(B) /R(0) = (R(B) - R(0)) /R(0)$,
\begin{eqnarray}
\frac{\triangle R(B)}{R(0)} & = & \frac{e^2}{\pi \hbar} \frac{R}{L} \Biggl\{ \frac{3}{2} \Biggl[ \Biggl(
\frac{1}{L_\varphi^2} + \frac{4}{3L_{so}^2} + \frac{1}{D \tau_B} \Biggr)^{-1/2} \nonumber \\
& - & \Biggl( \frac{1}{L_\varphi^2} + \frac{4}{3L_{so}^2}
\Biggr)^{-1/2} \Biggr] \nonumber \\
& - & \frac{1}{2} \Biggl[ \Biggl( \frac{1}{L_\varphi^2} + \frac{1}{D \tau_B}
\Biggr)^{-1/2} - L_\varphi \Biggr] \Biggr\} \,, \label{1D}
\end{eqnarray}
where $R$ is the resistance of a quasi-1D NW of length $L$, and the characteristic time scale $\tau_B$
represents the dephasing ability of the $B$ field. The form of $\tau_B$ depends on the orientation of
$B$ relative to the current flow and the shape of the NW \cite{Altshuler81}. There are two situations
which have been explicitly theoretically calculated. First, for a NW with a square cross section and
side $a$ in $B$ applied perpendicular to the NW axis, $\tau_B = 3L_B^4/(Da^2)$, where the magnetic
length $L_B = \sqrt{\hbar/eB}$. Second, for a NW with a circular cross section and diameter $d$ in $B$
applied parallel to the NW axis, $\tau_B = 8L_B^4/(Dd^2)$. In practice, both side $a$ and diameter $d$
can be treated as adjustable parameters, because the effective cross-sectional area responsible for
the charge conduction may differ from the geometric cross-sectional area of the given NW under study.
In other words, the NW may be inhomogeneous (e.g., due to O vacancies, surface absorption/desoption,
variations in compositions, surface states \cite{Shalish04}, accumulation layers
\cite{Grinshpan79,Gopel80}, etc.) and the electrical conduction is not through the whole volume of the
NW \cite{Schlenker08,Hu09}.

\begin{figure}[htp]
\begin{center}
\includegraphics[scale=0.23]{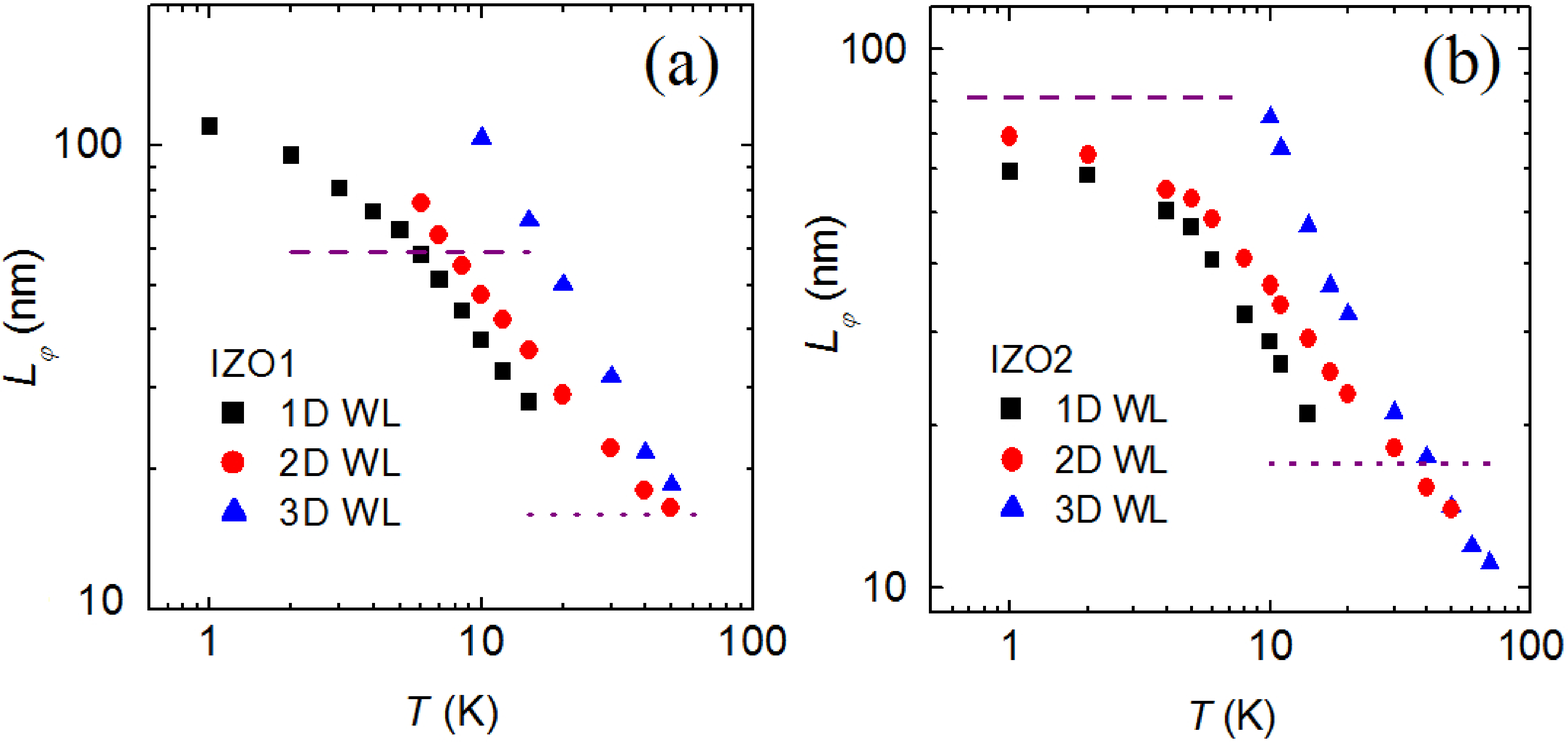}
\caption{Electron dephasing length $L_\varphi$ as a function of temperature for the (a) IZO1 and (b) IZO2 NWs.
Different symbols represent the $L_\varphi$ values extracted according to different dimensional WL MR expressions:
1D equation~(\ref{1D}) (squares), 2D equation~(\ref{2D}) (circles), and 3D  equation~(\ref{3D}) (triangles).
The dashed (dotted) line indicates the effective NW diameter $d$ (the outer conduction shell thickness $t$).
\label{fig4}}
\end{center}
\end{figure}

In plotting figure~\ref{fig3}(a), we have used equation~(\ref{1D}) and rewritten $\tau_B =
3L_B^4/(Da^2) \equiv 12L_B^4/(D\pi d^2)$ to least-squares fit the measured MR data for $B <$ 0.5 T,
and then generated the theoretical curves for a $B$ range up to 3.25 T. (We have treated $\tau_B$, and
thus $d$, as an adjustable parameter.) Numerically, we obtained $T$-independent (average) values of $d
\simeq$ 59 nm and $L_{so} \simeq$ 140 nm for all the MR curves plotted in figure~\ref{fig3}(a). The
only $T$-dependent adjustable parameter is $L_\varphi$, which is plotted in figure~\ref{fig4}(a) as a
function of $T$. Figure \ref{fig4}(a) shows that $L_\varphi$ (the squares) decreases from 110 nm at 1
K to 28 nm at 15 K. It should be noted that our extracted $L_\varphi$ value becomes shorter than the
effective NW diameter $d$ as $T$ increases to above $\sim$ 7 K. Such a result is not self-consistent
and it violates the applicability of equation~(\ref{1D}). That is, under such circumstances, one
should consider a possible dimensionality crossover to the 3D WL effect at $T >$ 7 K. A similar
observation has recently been pointed out by Hsu {\it et al.} \cite{Hsu-prb10} in their WL studies of
single ITO NWs. In any case, one should be cautious about the validity of the extracted $L_\varphi$
values in this moderately high $T$ region.

To examine whether a crossover from the 1D to the 3D WL effect takes place in the IZO1 NW at $T >$ 7
K, we now compare our MR data with the 3D WL theoretical predictions. The 3D WL MR is given by
\cite{Lin-jpcm02,Fukuyama81,Wu-prb94}, in terms of normalized magnetoresistivity $\triangle \rho (B)
/\rho^2(0) = (\rho (B) - \rho (0)) /\rho^2(0)$,
\begin{eqnarray}
\frac{\bigtriangleup \rho (B)}{\rho^2(0)} & = & \frac{e^2}{2 \pi^2 \hbar}\,
\sqrt{\frac{eB}{\hbar}}\, \Biggl\{ \frac{1}{2 \sqrt{1 - \gamma}}\, \biggl[\,
f_3 \Bigl( \frac{B}{B_-} \Bigr)
- f_3 \Bigl( \frac{B}{B_+} \Bigr)\, \biggr] \nonumber \\
& - & f_3 \Bigl( \frac{B}{B_2} \Bigr) - \sqrt{\frac{4B_{so}}{3B}} \nonumber \\
& \times & \biggl[\, \frac{1}{\sqrt{1 - \gamma}} \, ( \sqrt{t_+} - \sqrt{t_-}\,) +
\sqrt{t} - \sqrt{t + 1}\, \biggr] \Biggr\} \,,\,\, \label{3D}
\end{eqnarray}
where
$$
\gamma = \biggl[ \frac{3 g^\ast \mu_B B}{4eD ( 2B_{so} - B_0 ) }\, \biggr] ^2
\,,
$$
$$
t = \frac{3B_\varphi}{2 ( 2 B_{so} - B_0 )}\,, \,\, t_\pm = t + \frac{1}{2} (
1 \pm \sqrt{1 - \gamma}\, )\,,
$$
$$
B_\varphi = B_{in} + B_0 \,, \,\, B_2 = B_{in} + \frac{1}{3} B_0 + \frac{4}{3}
B_{so}\,,
$$
$$
B_\pm = B_\varphi + \frac{1}{3} ( 2B_{so} - B_0 ) ( 1 \pm
\sqrt{1 - \gamma} \, )\,,
$$
$g^\ast$ (= $-$1.93 in the ZnO material \cite{g-factor}) is the electron Lande-$g$ factor, $\mu_B$ is the Bohr magneton, and
\begin{eqnarray}
f_3(z) & \approx & 2 \biggl( \sqrt{2 + \frac1z} - \sqrt{\frac1z} \biggr) - \bigg[ \biggl( \frac12
+ \frac1z \biggr)^{-1/2} \nonumber \\ & + & \biggl( \frac32 + \frac1z \biggr)^{-1/2} \biggr] + \frac{1}{48} \biggl(
2.03 + \frac1z \biggr)^{-3/2} \,. \nonumber
\end{eqnarray}

Here the characteristic fields are connected with the electron scattering times through the relation
$B_j = \hbar/(4eD\tau_j)$, with the subscript $j$ stands for $\varphi$ (the dephasing field/time),
$in$ (the inelastic scattering field/time), $so$ (the s-o scattering field/time), $0$ (the
``saturated" scattering field/time as $T \rightarrow$ 0 K), and $e$ (the elastic scattering
field/time). ($B_e$ will be used in equation~(\ref{2D}).) The function $f_3$ is an infinite series
which can be approximately expressed as above, which is known to be accurate to be better than 0.1\%
for all arguments $z$ (Ref. \onlinecite{Baxter89}).

Figure \ref{fig3}(c) shows the normalized MR and the least-squares fits to the theoretical predictions
of equation~(\ref{3D}) at four $T$ values between 20 and 50 K. This figure reveals that, as $T$
increases to above $\sim$ 20 K, the theoretical curves can reasonably describe the experimental data.
The $L_\varphi$ values (triangles) thus extracted are plotted in figure~\ref{fig4}(a). We obtain
$L_\varphi$ = 50 nm at 20 K and 19 nm at 50 K. Although our measured MR data at $T >$ 20 K can
seemingly be described by the 3D WL theory, we should stress that the agreement between the theory and
the experiment is superficial. For instance, the inferred $L_\varphi$ values (triangles) do not
extrapolate to those $L_\varphi$ values (squares) inferred from the low-$T$ 1D regime. Furthermore,
the extracted $L_\varphi$ value would suggest a 3D-to-1D dimensionality crossover taking place at a
relatively high $T$ $\sim$ 20 K, which is very unlikely. (Recall that the least-squares fits to the 1D
MR theory do not lead to self-consistent results for $T \gtrsim$ 7 K.) In fact, as we will demonstrate
below, the charge carriers in our IZO NWs do not flow through the whole volume of the individual NWs.
Instead, there is an outer conduction shell of a thickness $t$ in individual IZO NWs
(figure~\ref{fig1}), which is responsible for our observed quantum-interference electron conduction.
Thus, at sufficiently low $T$ where $L_\varphi \gtrsim d$, the electrical transport would be 1D with
regard to the WL effect. On the other hand, at not too low $T$ where $L_\varphi \lesssim d$, the
low-field MR manifests the 2D WL effect. Note that, recently, \textit{surface-related electrical
conduction processes} have also been found in natively doped ZnO NWs which lie on the metallic side
of, but close to, the M-I transition \cite{Chiu-nano09ZnO}.

We now analyze our MR data at $T \gtrsim$ 7 K in terms of the 2D WL theory. The MR due to the 2D WL
effect in the presence of a perpendicular $B$ field is given by \cite{Bergmann84,Hikami80,Lin-prb87a},
in terms of normalized sheet resistance $\triangle R_\Box (B) /R_\Box^2(0) = (R_\Box (B) - R_\Box (0))
/R_\Box^2(0)$,
\begin{eqnarray}
\frac{\triangle R_\Box(B)}{R_\Box^2(0)} & = & \frac{e^2}{2\pi^2 \hbar} \biggl[
\Psi \biggl( \frac12 + \frac{B_1}{B} \biggr) - \frac32 \Psi \biggl( \frac12 +
\frac{B_2}{B} \biggr) \nonumber \\ & + & \frac12 \Psi \biggl( \frac12 + \frac{B_\varphi}{B}
\biggr) - \frac12 \ln \biggl( \frac{B_1^2 B_\varphi}{B_2^3} \biggr) \biggr]
\,\,, \label{2D}
\end{eqnarray}
where $\Psi$ is the digamma function, and $B_2$ was defined below equation~(\ref{3D}). $B_1$ is a
characteristic field given by $B_1 = B_e + B_{so} + B_0/2 \simeq B_e$. In the least-squares fits of
the predictions of equation~(\ref{2D}) to our experimental MR data, we have treated the sheet
resistance $R_\Box = R/(L/w)$ as an adjustable parameter, where the resistance $R$ and length $L$ are
directly measured, and $w$ is defined as $w = \pi (dia - t)$, with $dia$ being the geometric diameter
of the NW determined via SEM (table~\ref{t2}). Thus, $t$ is a fitting parameter. Note that we have
approximated the outer conduction shell with a dodecagon in the fits to equation~(\ref{2D})
\cite{note5}.

Figure \ref{fig3}(b) shows the measured MR and the least-squares fits to equation~(\ref{2D}) for the
IZO1 NW at several $T$ values between 7 and 50 K. This figure clearly indicates that the normalized MR
in the low-field regime of $B \lesssim$ 1 T can be well described by the theoretical predictions. Most
important, the extracted $L_\varphi$ values (circles) as a function of $T$ are plotted in
figure~\ref{fig4}(a), which lie systematically below those extracted according to the 3D form of
equation~(\ref{3D}). Inspection of figure~\ref{fig4}(a) indicates that the $L_\varphi$($\gtrsim$ 7 K)
values inferred from the 2D WL theory, equation~(\ref{2D}), {\em closely extrapolate to} those
$L_\varphi$($\lesssim$ 7 K) values inferred from the 1D WL theory, equation~(\ref{1D}). This
observation is meaningful, which strongly suggests that the WL MR effect {\em smoothly crosses over}
from the 1D regime to the 2D regime as $T$ increases to be above $\approx$ 7 K. Indeed, we find that
the measured MR data at 7 K and in $B \lesssim$ 1 T can be reasonably well described by both
equation~(\ref{1D}) and equation~(\ref{2D}) (figure~\ref{fig3}(d)). Furthermore, it should be noted
that, at this particular $T$ value, the fitted dephasing length according to equation~(\ref{2D}) is
$L_\varphi$(7\,K) $\simeq$ 64 nm. This is very close to the effective NW diameter $d$ ($\simeq$ 59 nm)
inferred above from the 1D WL fits. Thus, the physical quantity $d$ signifies the characteristic
length scale that controls the dimensionality crossover between the 1D ($L_\varphi \gtrsim d$) and 2D
($L_\varphi \lesssim d$) WL regimes in the IZO1 NW.

If $T$ continues to increase and $L_\varphi$ gradually reduces, one would expect another possible
dimensionality crossover from the 2D to the 3D WL effect. In fact, figures \ref{fig3}(b) and
\ref{fig3}(c) together indicate that both equation~(\ref{3D}) and equation~(\ref{2D}) can describe the
measured MR data at 50 K and in $B \lesssim$ 1.5 T. The extracted $L_\varphi$(50 K) values according
to both equations approach each other, being $\approx$ 15 nm (figure~\ref{fig4}(a)). This result
implies that a 2D-to-3D dimensionality crossover takes place around $T$ $\sim$ 50 K or slightly
higher. In particular, the responsible length scale is $\approx$ 15 nm, which can be identified as the
effective thickness $t$ of the outer conduction shell. In short, in the IZO1 NW, as $T$ monotonically
increases and $L_\varphi$ progressively decreases, a 1D-to-2D dimensionality crossover of the WL
effect first takes place around $\sim$ 7 K, where $L_\varphi$(7 K) $\simeq$ $d$. As $T$ further
increases, a 2D-to-3D dimensionality crossover eventually occurs near $\sim$ 50 K, where
$L_\varphi$(50 K) $\simeq$ $t$. It should be noted that the existence of a relevant shell thickness of
$\approx$ 15 nm is further supported by the EEI effect in the $R$-$T$ behavior (figures~\ref{fig6}(a)
and \ref{fig6}(b)). Thus, continuously fitting the measured MR curves with the 2D WL theory up to $T
>$ 50 K would lead to an inconsistency of $L_\varphi^{\rm (2D)} < t$.

Apart from the IZO1 NW, we have carried out similar MR measurements on the IZO2 NW. In this second NW,
we obtain least-squares fitted average values of $d$ $\simeq$ 81 nm, $t$ $\simeq$ 17 nm, and $L_{so}$
$\simeq$ 105 nm. Figure \ref{fig4}(b) shows the $L_\varphi$ values extracted according to the 1D, 2D,
and 3D WL MR expressions, as indicated. We find that it definitely needs to apply the 2D WL theory to
extract acceptable values of $L_\varphi$ (circles) in this NW. Those $L_\varphi$ values (squares)
extracted according to equation~(\ref{1D}) are smaller than $d$ at all $T$, and thus an interpretation
based on the 1D WL effect is not self-consistent. On the other hand, those $L_\varphi$ values
(triangles) extracted according to equation~(\ref{3D}) are larger than $t$ at $T$ below $\sim$ 40 K,
and thus the 3D WL effect is neither acceptable for $T$ $<$ 40 K. In other words, our results suggest
that the IZO2 NW lies in the 2D regime with regard to the WL effect all the way down to $T$ $\sim$ 1
K. There is no crossover to the 1D WL regime because this NW has a larger $d$ value while it possesses
a shorter $L_\varphi$, as compared with the IZO1 NW. Physically, the $L_\varphi$ is shorter in the
IZO2 NW because this NW is slightly more disordered than the IZO1 NW. In the opposite high $T$ region,
there is a 2D-to-3D dimensionality crossover taking place around 40 K. At 40 K, the measured MR can be
fitted with both equation~(\ref{3D}) and equation~(\ref{2D}), and the extracted values are similar,
being $L_\varphi$(40 K) $\equiv$ $t$ $\simeq$ 17 nm. It should be noted that this $t$ value is
relatively close to that ($\simeq$ 15 nm) inferred for the IZO1 NW. These results suggest that this is
the typical thickness of the surface conduction shell in our IZO NWs. This characteristic thickness
might be a material property of the group-III metal doped ZnO NWs \cite{Look08}.

\subsection{Two-dimensional parallel magnetoresistance in the weak-localization effect}

In this subsection, we intend to provide further evidence for the existence of a surface conduction
shell in individual IZO NWs. If there is any quasi-2D structure leading to the 2D WL effect observed
in our IZO NWs discussed thus far, measuring the MR in $B$ applied parallel to the conduction shell
should provide a complementary method for determining the thickness $t$ of this conduction shell. The
MR due to the 1D WL effect in the presence of a parallel magnetic field, $B_{||}$, is already included
in equation~(\ref{1D}). The MR due to the 2D WL effect in the presence of a parallel $B_{||}$ is given
by \cite{ng93}, in terms of normalized sheet resistance $\triangle R_\Box (B_\|)/R_\Box^2(0) = (R_\Box
(B_\|) - R_\Box (0)) /R_\Box^2(0)$,
\begin{equation}
\frac{\triangle R_\Box(B_\|)}{R_\Box^2(0)} = - \frac{e^2}{2\pi^2 \hbar}
\biggl[ \frac32 \ln \biggl( 1 + \frac{L_2^2}{L_\|^2} \biggr) - \frac12 \ln
\biggl( 1 + \frac{L_\varphi^2}{L_\|^2} \biggr) \biggr] \,\,, \label{2Dp}
\end{equation}
where the characteristic lengths $L_\| = \sqrt{3}\hbar /(eB_\| t)$, and $L_2 =
\sqrt{\hbar/(4eB_2)}$, with the characteristic fields $B_2$ and $B_\varphi$
being defined below equation~(\ref{3D}).

\begin{figure}[htp]
\begin{center}
\includegraphics[scale=0.23]{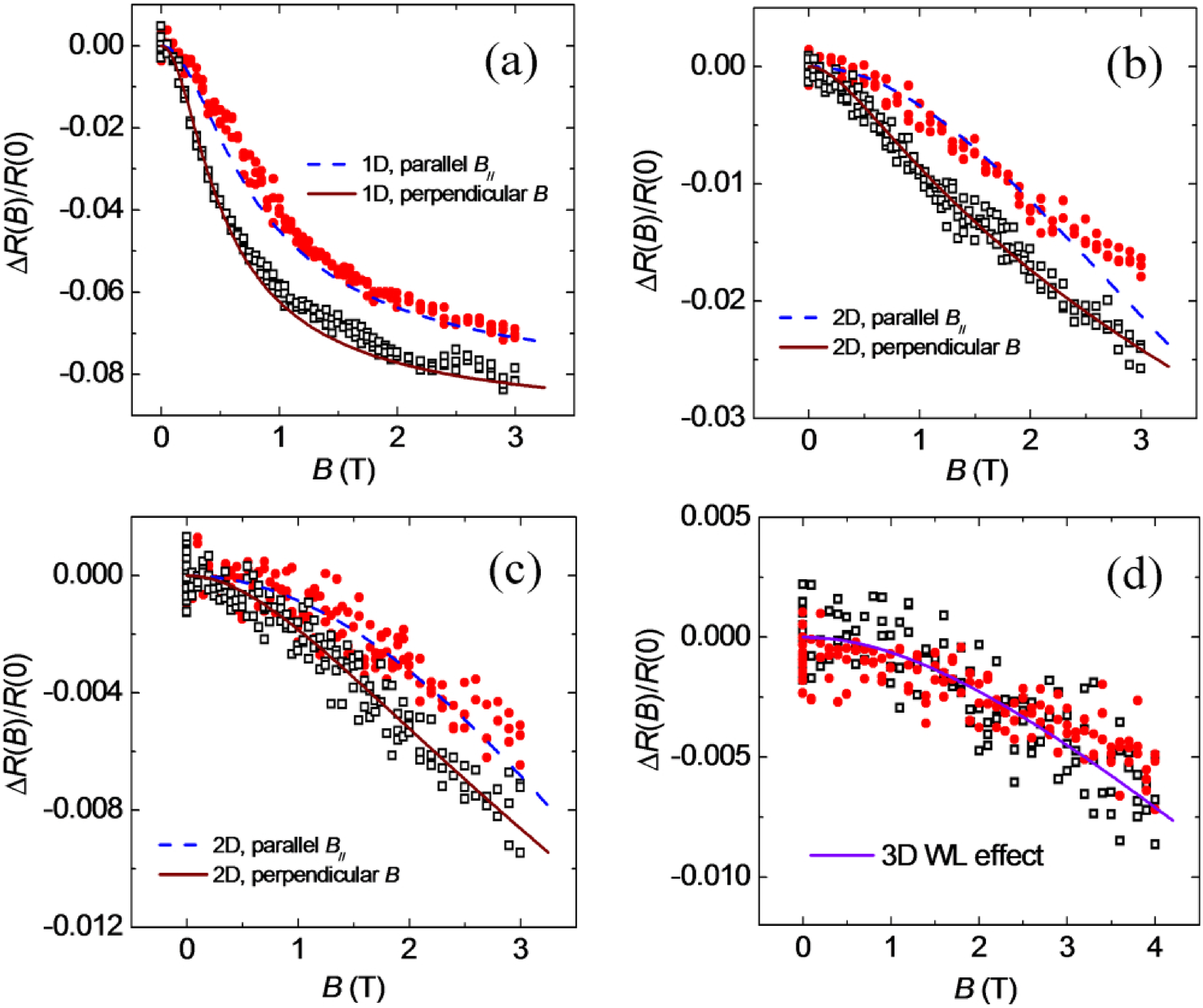}
\caption{Normalized MR, $\triangle R(B)/R(0) = (R(B) - R(0))/R(0)$, as a function of magnetic field for
the IZO1 NW at (a) 3.0 K, (b) 20 K, and (c) 50 K. (d) $\triangle R(B)/R(0)$ as a function of magnetic
field of the IZO2 NW at 70 K. In each figure, open squares (closed circles) are the perpendicular
(parallel) MR data. The solid (dashed) curves are the theoretical predictions of the perpendicular
(parallel) WL effects. The solid curve in (d) is a least-squares fit to equation~(\ref{3D}) with
$L_\varphi$ = 11 nm. \label{fig5}}
\end{center}
\end{figure}

Figures \ref{fig5}(a), \ref{fig5}(b), and \ref{fig5}(c) plot the perpendicular and the parallel MR
data of the IZO1 NW at three selected $T$ values of 3 K (i.e., the 1D WL regime), 20 K (i.e., the 2D
WL regime), and 50 K (i.e., the 2D-to-3D crossover regime), respectively. In each figure, the solid
(dashed) curve is the theoretical prediction of the perpendicular (parallel) WL effect. We start with
the low-$T$ 1D regime. In figure~\ref{fig5}(a), the perpendicular MR was first least-squares fitted to
equation~(\ref{1D}) with the form of $\tau_B = 12L_B^4/(D \pi d^2)$ as discussed in subsection 3.2,
and the solid curve was plotted using the fitted parameters: $L_\varphi$(3 K) = 81 nm, $L_{so}$ = 140
nm, and $d$ = 59 nm. Then, we substituted this set of parameters into equation~(\ref{1D}) but now
rewrote the magnetic time in the form of $\tau_B = 8L_B^4/(D d^2)$ to ``generate" the 1D {\em
parallel} WL MR prediction, {\em without} invoking any additional adjustable parameter nor performing
any further least-squares fits. This procedure produced the dashed curve which is seen to well
describe our measured parallel MR data. Thus, the reliability and validity of our measurement method
and data analyses are justified. In fact, we could repeat this practice of using the {\em same} set of
fitting parameters to describe both the perpendicular and the parallel MR data at a given $T$ for all
temperatures below $\approx$ 7 K. In figure~\ref{fig5}(a), the magnitudes of the perpendicular MR and
the parallel MR do not differ markedly, i.e., the MR is not significantly anisotropic, because
$L_\varphi$(3 K) is not considerably longer than $d$ in this NW \cite{note6}.

At $T$ $>$ 7 K, our measured perpendicular MR data and the parallel MR data at a given $T$ can {\em no
longer} be simultaneously described by equation~(\ref{1D}) with a same set of adjustable parameters.
Therefore, we have turned to the 2D forms of the WL theory. In this procedure, if we already know the
$L_\varphi$ and $L_{so}$ values from the analyses of the perpendicular MR data at a given $T$ as
discussed in subsection 3.2, the conduction shell thickness $t$ (which enters $L_\|$) will be the
\textit{sole} adjustable parameter left in equation~(\ref{2Dp}). Figures \ref{fig5}(b) and
\ref{fig5}(c) clearly show that our perpendicular MR data and parallel MR data can be simultaneously
described by equation~(\ref{2D}) (solid curves) and equation~(\ref{2Dp}) (dashed curves),
respectively. From this approach, our extracted $t$ value according to equation~(\ref{2Dp}) is
$\simeq$ 15 nm at both temperatures, firmly confirming the above deduced thickness. In fact, by
repeating for several $T$ values in the 2D WL regime, we obtain average values of $t \simeq$ 15$\pm$2
nm for the IZO1 NW and $t \simeq$ 17$\pm$3 nm for the IZO2 NW (table~\ref{t3}).

Figure \ref{fig5}(d) shows a plot of the perpendicular MR data (open squares) and the parallel MR data
(closed circles) as a function of $B$ for the IZO2 NW at 70 K. One sees that the perpendicular MR and
the parallel MR collapse, explicitly illustrating a 3D behavior. Indeed, at such a high $T$ value,
$L_\varphi$ must be very short. According to equation~(\ref{3D}) (the solid curve), we obtain a
least-squares fitted value of $L_\varphi$(70 K) $\simeq$ 11 nm.

\textit{Self purification mechanisms preventing doping of semiconductor nanostructures.} It should be
of crucial importance to point out that, from the Hall effect and secondary-ion mass spectroscopy
measurements, Look {\it et al.} \cite{Look08} have recently inferred that the group-III metal
impurities could readily diffuse into the surfaces of any ZnO wafers for a distance of $\approx$ 14
nm. Remarkably, this value independently inferred from entirely distinct physical properties is in
close agreement with our extracted $t$ value ($\simeq$ 15--17 nm). The underlying physics for this
consistency is highly meaningful. Recently, based on energetic arguments, Dalpian and Chelikowsky
\cite{Dalpian06} have theoretically shown that the ``self-purification" mechanisms would perniciously
prevent doping of semiconductor nanostructures, causing dopants to segregate to surfaces. This
theoretical finding provides a natural explanation for our experimental observation of the
``core-shell-like structure" in IZO NWs. This very issue concerning the materials property and the
doping behavior of semiconductors at the nanoscale deserves detailed studies before any nanoelectronic
devices could be possibly implemented \cite{Klamchuen11}.

\textit{The absence of Altshuler-Aronov-Spivak (AAS) oscillations.} It may be conjectured that a
convincing experimental proof of the existence of an outer conduction shell would be an observation of
the AAS oscillations at low temperatures \cite{AAS82}. AAS had theoretically predicted that the
resistance of a weakly disordered cylindrical conductor would oscillate in sweeping $B_\|$ fields with
a period of $h/2e$. We have checked this predicted phenomenon in this study, but did not observe any
signature of such kind of oscillations. This is expected, because a conduction shell of as thick as
$\simeq$ 15--17 nm in our IZO NWs would strongly suppress the amplitudes of the AAS oscillations,
making them more than one order of magnitudes smaller than the parallel MR in the WL effect
\cite{note7}. Moreover, the AAS oscillations should be further damped due to any inhomogeneities in
the conduction shell radius and thickness \cite{Aronov87}, which very likely exist in our NWs.

\begin{table*}
\caption{\label{t3} Least-squares fitted parameters for IZO NWs. $A_{ee}^{\rm (1D)}$, $A_{ee}^{\rm
(2D)}$, and $A_{ee}^{\rm (3D)}$ are the electron-electron scattering strength in the 1D, 2D, and 3D
regimes, respectively. $\tau_0$ is the electron dephasing time as $T$ $\rightarrow$ 0 K, $\tau_{so}$
is the spin-orbit scattering time, $d$ is the effective NW diameter, $w$ is the perimeter of the
dodecagon assumed in the least-squares fits to equation~(\ref{2D}), and $t$ is the outer conduction
shell thickness.}
\begin{ruledtabular}
\begin{tabular}{lcccccccc}
Sample & $A_{ee}^{\rm (1D)}$ & $A_{ee}^{\rm (2D)}$ & $A_{ee}^{\rm
(3D)}$ & $\tau_0$ & $\tau_{so}$ & $d$ & $w$ & $t$ \\

 & K$^{-2/3}$\,s$^{-1}$ & K$^{-1}$\,s$^{-1}$ & K$^{-3/2}$\,s$^{-1}$
 & (ps) & (ps) & (nm) & (nm) & (nm) \\

\hline

IZO1 & 1.6$\times$10$^{10}$ & 1.1$\times$10$^{10}$ & 2.3$\times$10$^9$
& $\sim$ 67 & 54 & 59 & 159 & 15$\pm$2 \\
IZO2 & --- & 9.0$\times$10$^9$ & 2.2$\times$10$^9$ & $\sim$ 40 & 47
& 81 & 235 & 17$\pm$3 \\

\end{tabular}
\end{ruledtabular}
\end{table*}

\section{Dimensionality crossover in the electron-electron interaction effect}

In this section, we concentrate on the $R(T)$ data due to the EEI effect to provide further
justification for the existence of a core-shell-like structure in IZO NWs. In addition to the WL
effect, the many-body EEI effect also results in a resistance rise with decreasing $T$ in a weakly
disordered conductor \cite{Bergmann84,Bergmann10,Altshuler85}. The EEI effect induced correction in 2D
is give by \cite{Altshuler85,Lin-prb87a}, in terms of the normalized sheet resistance $\triangle
R_\Box(T)/R_\Box(T_0) = (R_\Box(T) - R_\Box(T_0))/R_\Box(T_0)$,
\begin{equation}
\frac{\triangle R_\Box (T)}{R_\Box (T_0)} = - \frac{e^2}{2 \pi^2 \hbar}
\biggl( 1 - \frac34 \tilde{F} \biggr) R_\Box \ln \biggl( \frac{T}{T_0} \biggr)
\,, \label{2Dee}
\end{equation}
where $T_0$ is an arbitrary reference temperature. $\tilde{F}$ is an electron screening factor
averaged over the Fermi surface, whose value lies approximately between 0 and 1
\cite{Altshuler85,F-value}. The EEI effect induced correction in 3D is give by
\cite{Altshuler85,Lin-prb93}, in terms of the normalized resistivity $\triangle \rho(T)/\rho(T_0) =
(\rho(T) - \rho(T_0)) /\rho(T_0)$,
\begin{equation}
\frac{\triangle \rho (T)}{\rho (T_0)} = - \frac{0.915 e^2}{4 \pi^2 \hbar}
\biggl( \frac43 - \frac32 \tilde{F} \biggr) \rho \, \sqrt{\frac{k_B}{\hbar D}}
\, \biggl(\sqrt{T} - \sqrt{T_0} \biggr) \,. \label{3Dee}
\end{equation}
In the comparison of the theoretical prediction of either equation~(\ref{2Dee}) or
equation~(\ref{3Dee}) with experiment, $\tilde{F}$ is the only adjustable parameter. The
characteristic length scale controlling the sample dimensionality in the EEI effect is the thermal
diffusion length $L_T$ = $\sqrt{D\hbar /k_BT}$. Note that the EEI effect in different sample
dimensionalities results in distinct $T$ dependencies of the resistance rise at low temperatures
\cite{note8}. The WL effect induced corrections to the residual resistance at low $T$ can be readily
suppressed by applying a moderately high $B$ field \cite{Bergmann84,Bergmann10,Altshuler85}.
Therefore, one may measure $R(T)$ in an applied field to focus on the EEI term alone.

\begin{figure}[htp]
\begin{center}
\includegraphics[scale=0.23]{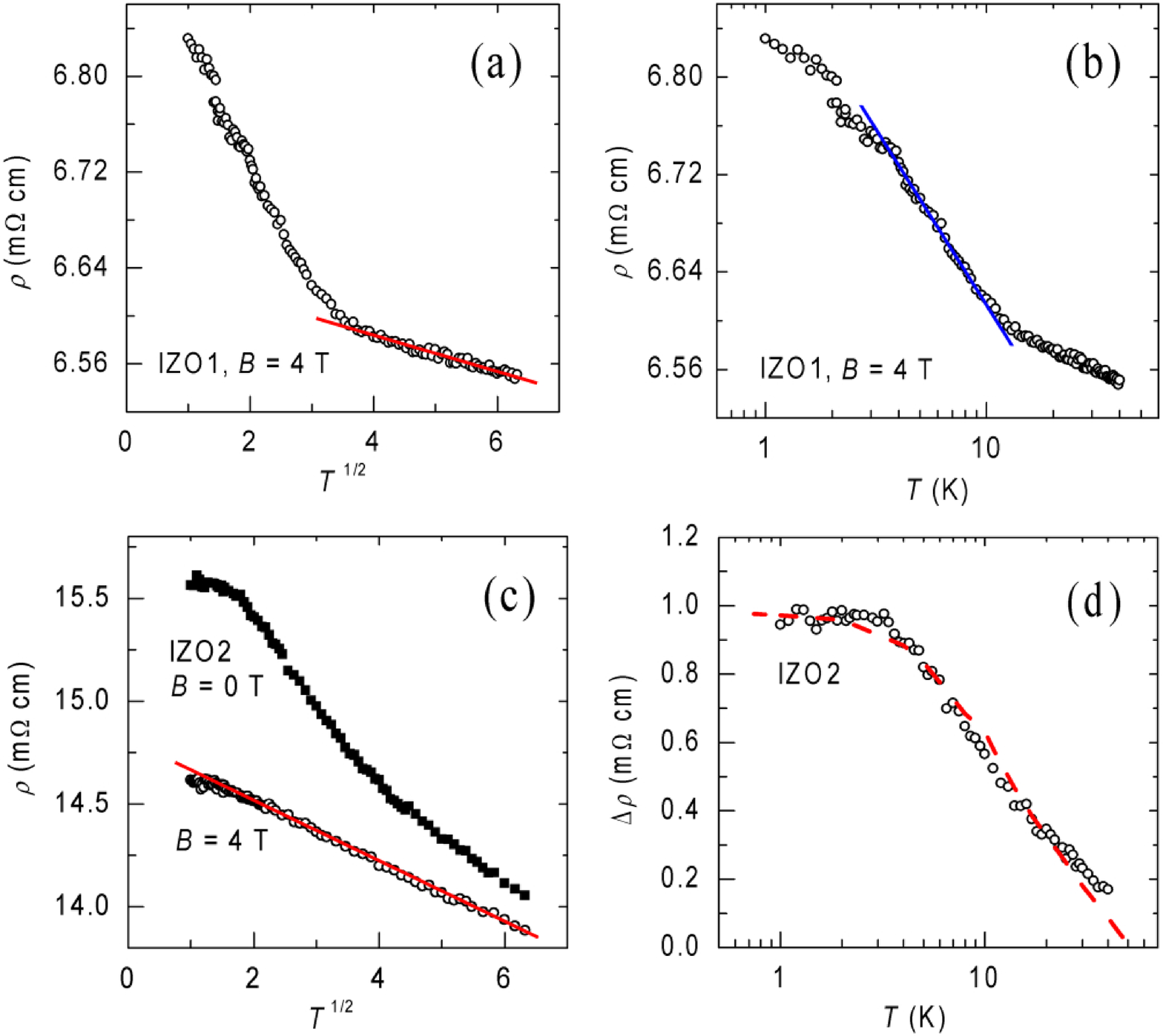}
\caption{Resistivity of the IZO1 NW measured in a perpendicular $B$ field of 4 T as a function of
(a) the square root of temperature, and (b) the logarithm of temperature. (c) Resistivity as a
functions of the square root of temperature for the IZO2 NW measured in $B = 0$ and in perpendicular
$B$ = 4 T, as indicated. (d) A plot of the difference in resistivity from (c),
$\triangle \rho = \rho (B\,=\,0) - \rho (B\,=\,4 \,\,{\rm T})$, as a function of the logarithm of
temperature. The straight solid line in (b) is a least-squares fit to equation~(\ref{2Dee}), and the
straight solid lines in (a) and (c) are least-squares fits to equation~(\ref{3Dee}). The dashed curve
in (d) is the theoretical prediction of equation~(\ref{2DWL}). \label{fig6}}
\end{center}
\end{figure}

We start with our discussion on the IZO2 NW. Figure \ref{fig6}(c) shows the resistivity of the IZO2 NW
as a function of $\sqrt{T}$ in both $B$ = 0 and in a perpendicular $B$ = 4 T, as indicated. Clearly,
the 4-T data illustrate a robust linear dependence between $\sim$ 3 and 40 K. Such a $-$$\sqrt{T}$
temperature dependence manifests the 3D EEI effect in this wide $T$ interval. By comparing with the
prediction of equation~(\ref{3Dee}), we obtained a value $\tilde{F}$ $\simeq$ 0.42. This magnitude of
$\tilde{F}$ is in line with that found in typical doped semiconductors, such as Si:B \cite{Dai92}. On
the other hand, in $B$ = 0, the $T$ dependence of $\rho$ is somewhat more complicated, because both
the WL and the EEI effects now contribute to the total resistivity rise. Using $D$ = 2.2 cm$^2$/s
(table \ref{t2}), we estimate $L_T$(5 K) $\simeq$ 18 nm. This length scale is basically the conduction
shell thickness inferred from the WL MR studies (section 3). That is, the 3D EEI effect on the
resistivity rise is expected to persist from intermediately high $T$ down to $\sim$ 5 K in this
particular NW. This prediction is in good accord with the observation depicted in
figure~\ref{fig6}(c).

Figure \ref{fig6}(d) plots the variation of the difference in the measured resistivity, $\triangle
\rho$ = $\rho$($B$\,=\,0) $-$ $\rho$($B$\,=\,4 T), with temperature of the IZO2 NW whose resistivities
are shown in figure~\ref{fig6}(c). This figure indicates an approximate ln\,$T$ temperature dependence
of $\triangle \rho$ between $\sim$ 3 and $\sim$ 40 K. This observation is meaningful. Indeed, this
$\triangle \rho$ can be identified as originating from the 2D WL effect, which is theoretically
predicted to be given by \cite{Altshuler85}, in terms of the normalized sheet resistance,
\begin{equation}
\frac{\triangle R_\Box (T)}{R_\Box (T_0)} = \frac{e^2}{4 \pi^2 \hbar} R_\Box
\biggl\{ \ln \biggl[ \frac{B_\varphi (T)}{B_\varphi (T_0)} \biggr] - 3\, \ln
\biggl[ \frac{B_2 (T)}{B_2 (T_0)} \biggr] \biggl\} \,, \label{2DWL}
\end{equation}
where $T_0$ is an arbitrary reference temperature, and $B_2$ and $B_\varphi$ are defined below
equation~(\ref{3D}). Writing $\rho$ = $t$$R_\Box$ and substituting the fitted values of $L_\varphi$
and $L_{so}$ from our MR data analyses described in section 3 into equation~(\ref{2DWL}), we obtain
the dashed curve shown in figure~\ref{fig6}(d), which is seen to satisfactorily describe the
experimental $\triangle \rho$. This result strongly confirms our scenario of the quantum-interference
transport through a surface layer in single IZO NWs. Recall that the 2D WL MR effect in the IZO2 NW
has been observed in the same $T$ interval of 1--40 K (subsection 3.2). Below about 3 K, $\triangle
\rho$ tends to saturate to a constant value, because $\tau_\varphi$ becomes very weakly dependent on
$T$, as shown in figure~\ref{fig4}(b).

We turn to the IZO1 NW. For simplicity, we shall present and discuss only the $\rho (T)$ data measured
in a perpendicular $B$ = 4 T. Figure \ref{fig6}(a) shows that the $\rho \propto$ $-$$\sqrt{T}$ law
holds between $\sim$ 14 and 40 K. By comparing with equation~(\ref{3Dee}), we obtained a value of
$\tilde{F}$ $\simeq$ 0.61. In contrast to the case of the IZO2 NW, a change in the $T$ dependence to
the 2D $R_\Box$ $\propto$ $-$ln$T$ law is seen between $\sim$ 3.6 and $\sim$ 11 K in this NW, see
figure~\ref{fig6}(b). By writing $R_\Box$ = $\rho /t$ and comparing with the prediction of
equation~(\ref{2Dee}), we obtained a value $\tilde{F}$ $\simeq$ 0.51. This value is reasonably in line
with that inferred above from the high-$T$ 3D regime. Thus, a dimensionality crossover with regard to
the EEI effect does occur in the $\sim$ 11--14 K temperature window in this particular NW. In fact,
using $D$ = 3.6 cm$^2$/s (table \ref{t2}), we estimate $L_T$(11 K) = 16 nm. This value is very close
to the conduction shell thickness $t$ $\simeq$ 15 nm inferred from the WL MR studies discussed in
section 3. In other words, at $T$ $\gtrsim$ 11 K, $L_T$ $\lesssim$ $t$ and the EEI effect is 3D
(figure~\ref{fig6}(a)), while at $T$ $\lesssim$ 11 K, $L_T$ $\gtrsim$ $t$ and the EEI effect is 2D
(figure~\ref{fig6}(b)). In short, the observations of a 2D-to-3D dimensionality crossover in both the
EEI effect and the WL effect strongly substantiate the existence of an outer conduction shell of a
thickness $t$ in IZO NWs.

Finally, it may be readily estimated that the thermal lengths $L_T (T)$ = 52/$\sqrt{T}$ nm in the IZO1
NW and 41/$\sqrt{T}$ nm in the IZO2 NW. These lengths are relatively short, as compared with $d$.
Thus, a dimensionality crossover of the EEI effect from the 2D to the 1D regime is not seen in this
experiment. Table \ref{t1} summarizes the various $T$ intervals over which different dimensionalities
in the WL and the EEI effects are observed in this work.

\section{Electron dephasing time}

\begin{figure}[htp]
\begin{center}
\includegraphics[scale=0.20]{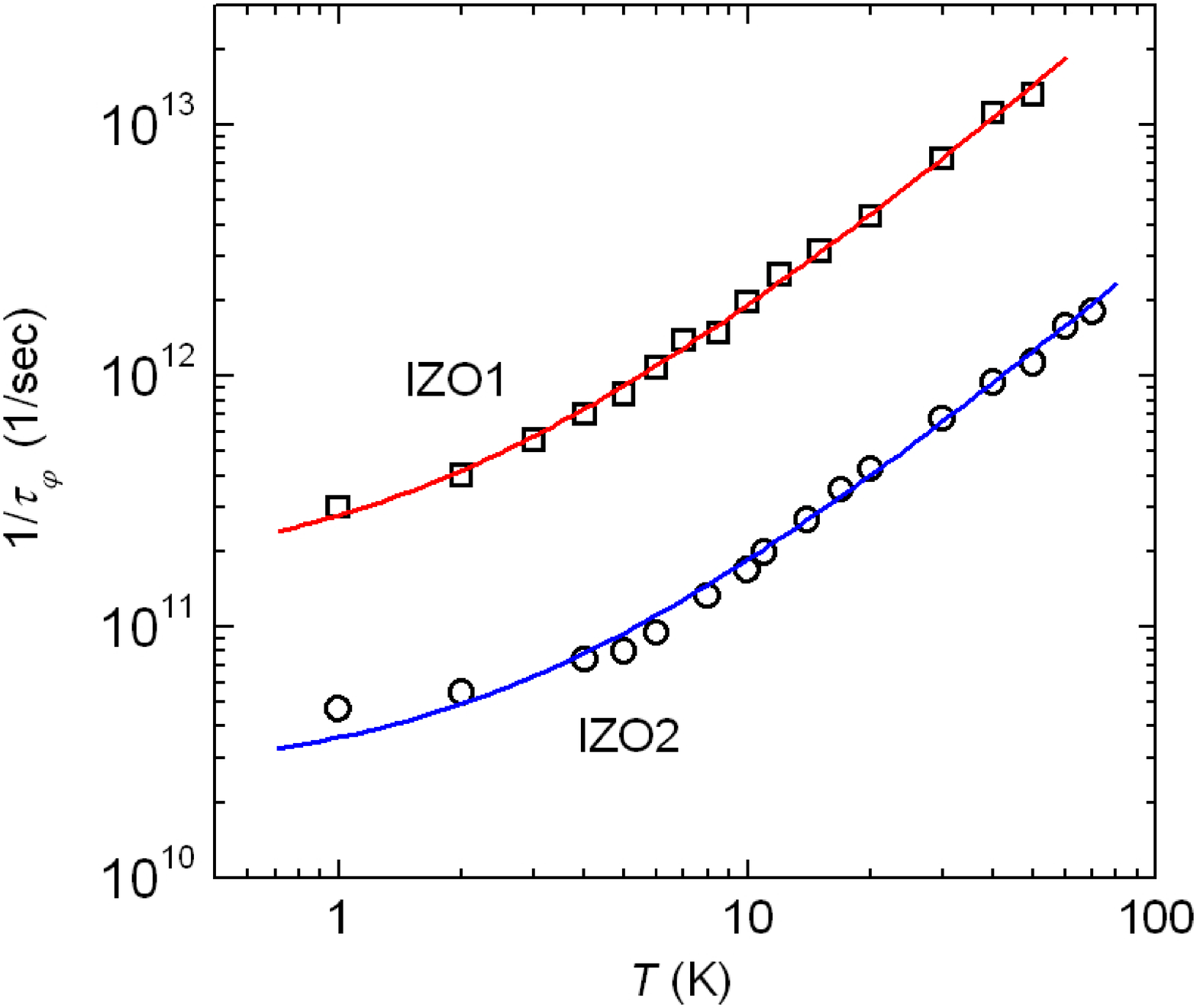}
\caption{Electron dephasing rate $\tau_\varphi^{-1}$ as a function of temperature for the IZO1 and
IZO2 NWs, as indicated. The solid curves are the theoretical predictions of equation~(\ref{rate}).
For clarity, the data of the IZO1 NW have been shifted up by multiplying a factor of 10. Note that
the two data points at 1 K might be subject to slight electron heating. \label{fig7}}
\end{center}
\end{figure}

In this section, we analyze the $T$ dependence of $\tau_\varphi^{-1}$ to study the electron dephasing
processes in IZO NWs. Recall that, for a given IZO NW, the ``correct" $\tau_\varphi^{-1}$ values in
different $T$ regions are those extracted according to the appropriate WL MR expressions in different
dimensionalities. Figure \ref{fig7} plots our extracted $\tau_\varphi^{-1}$ as a function of $T$ for
the IZO1 and IZO2 NWs, as indicated. In this figure, the $\tau_\varphi^{-1}$ values of the IZO1 NW are
a combination of those extracted according to the 1D WL form (equation~(\ref{1D})) between 1 and 8.5 K
and those extracted according to the 2D WL form (equation~(\ref{2D})) between 7 and 50 K. The
$\tau_\varphi^{-1}$ values of the IZO2 NW are a combination of those extracted according to the 2D WL
form (equation~(\ref{2D})) between 1 and 40 K and those extracted according to the 3D WL form
(equation~(\ref{3D})) between 40 and 70 K.

The physical meaning of $\tau_\varphi^{-1}$ is examined in the following. The total electron dephasing
rate in a weakly disordered degenerate semiconductor can be written as \cite{Lin-jpcm02}
\begin{equation}
\frac{1}{\tau_\varphi (T)} = \frac{1}{\tau_0} + \frac{1}{\tau^N_{ee}(T)} +
\frac{1}{\tau_{in}(T)} \,, \label{rate}
\end{equation}
where $\tau_0$  is a constant or a very weakly $T$ dependent quantity, whose origins (paramagnetic
impurity scattering, dynamical structural defects, etc.) are a subject of elaborate investigations in
the past three decades \cite{Birge03,Lin-prb87b,Mohanty97,Huang-prl07}. The quasi-elastic (i.e.,
small-energy-transfer) Nyquist electron-electron ($e$-$e$) relaxation rate, $(\tau^N_{ee})^{-1}$, in
low-dimensional disordered conductors is known to dominate $\tau_\varphi^{-1}$ in an appreciable $T$
interval. In the following analyses, we use the standard expression: $(\tau^N_{ee})^{-1}$ =
$A_{ee}T^p$, where $p$ = 2/3 and 1 for 1D and 2D samples, respectively
\cite{Altshuler85,Lin-jpcm02,Altshuler82}. Note that, as $T$ increases, the exponent of temperature
$p$ in $(\tau^N_{ee})^{-1}$ is expected to change as the sample dimensionality changes in our IZO NWs.
The third term, $\tau_{in}^{-1}$, on the right-hand side of equation~(\ref{rate}) denotes any
additional inelastic scattering mechanism(s) that might play a role in the dephasing process at
sufficiently high $T$.

Before comparing our experimental $\tau_\varphi^{-1}$ data with equation~(\ref{rate}), we first would
like to comment on the electron-phonon ($e$-ph) relaxation process. In disordered metals, the $e$-ph
scattering is often significant at a few degrees of kelvin and higher. One may then safely identify
the $\tau_{in}^{-1}$ term in equation~(\ref{rate}) as the $e$-ph scattering rate $\tau_{e-ph}^{-1}$
\cite{Sergeev-prb00}, in either the diffusive limit \cite{Lin-epl95,Zhong-prl98} or the
quasi-ballistic limit \cite{Zhong-prl10}, depending on the experimental conditions. However, the
carrier concentrations $n$ $\sim$ 1$\times$$10^{19}$ cm$^{-3}$ (table \ref{t2}) in our IZO NWs, which
are 3 to 4 orders of magnitude lower than those in typical metals \cite{Kittel}.  Theoretical
evaluations show that at such carrier concentrations, the deformation potential still has a metallic
nature, but substantially decreases due to the low concentration. The corresponding $e$-ph coupling
constant (the constant $\beta$ in \cite{Sergeev-prb00,Zhong-prl10}) is found to be proportional to
$n$. Therefore, the $e$-ph relaxation must be {\em negligible} in this work \cite{note9}. On the
contrary, it should be pointed out that the $e$-$e$ scattering in the IZO material relative to that in
typical metals is enhanced, owing to the smaller $E_F$ value and the larger $\rho$ ($R_\Box$) value in
IZO NWs.

In 3D, it is established that the $e$-$e$ scattering is determined by the large-energy-transfer
processes. In the clean limit, the theory \cite{Altshuler85,Lin-jpcm02} predicts
$(\tau_{ee}^{-1})_{\rm clean}$ = $\pi (k_BT)^2/(8 \hbar E_F)$. Substituting the $E_F$ values of our
IZO NWs into this expression, we estimate this scattering rate to be $\approx$ 4$\times$$10^7$ $T^2$
s$^{-1}$ ($\approx$ 8$\times$$10^7$ $T^2$ s$^{-1}$) in the IZO1 (IZO2) NW. Even at a moderately high
$T$ of 40 K, this scattering rate is more than (about) one order of magnitude smaller than the
experimental $\tau_\varphi^{-1}$ value in the IZO1 (IZO2) NW. Therefore, this clean-limit $e$-$e$
scattering process can be ruled out in the present study. In the dirty limit, the
large-energy-transfer $e$-$e$ scattering rate is modified to be $\tau_{ee}^{-1}$ = $A_{ee}^{\rm
(3D,th)}T^{3/2}$, with the coupling strength given by \cite{Altshuler85,Lin-jpcm02,Altshuler82}
\begin{equation}
A_{ee}^{{\rm (3D,th)}}= \frac{\sqrt{3}}{2 \hbar \sqrt{E_F}} \biggl(
\frac{k_B}{k_F \ell} \biggr)^{3/2} \,. \label{rate3D}
\end{equation}
This scattering rate is relevant to our experiment at high $T$ values where our NWs enter the 3D WL
regime.

Since the dimensionality crossover in the WL effect is less complex in the IZO2 NW, we first analyze
the $\tau_\varphi^{-1}$ data in this sample. In this NW, there is a single 2D-to-3D crossover in the
wide $T$ interval of 1--70 K. Therefore, we may rewrite equation~(\ref{rate}) in the following form:
$\tau_\varphi^{-1}$ = $\tau_0^{-1}$ + $A_{ee}^{{\rm (2D)}}T$ + $A_{ee}^{{\rm (3D)}}T^{3/2}$, where
$A_{ee}^{{\rm (2D)}}$ and $A_{ee}^{{\rm (3D)}}$ denote the $e$-$e$ scattering strength in the 2D and
3D regimes, respectively. (One may identify $A_{ee}^{{\rm (3D)}}T^{3/2}$ as the $\tau_{in}^{-1}$ term
in equation~(\ref{rate}).) By least-squares fitting this expression to the experimental data (see
figure~\ref{fig7}), we obtain $A_{ee}^{{\rm (2D)}}$ $\simeq$ 9.0$\times$$10^9$ K$^{-1}$ s$^{-1}$ and
$A_{ee}^{{\rm (3D)}}$ $\simeq$ 2.2$\times$$10^9$ K$^{-3/2}$ s$^{-1}$. Theoretically, the Nyquist
$e$-$e$ scattering strength in 2D is given by \cite{Altshuler85,Lin-jpcm02,Altshuler82}
\begin{equation}
A_{ee}^{{\rm (2D,th)}}= \frac{e^2 k_B}{2 \pi \hbar^2} \, R_\Box \, {\rm ln}
\biggl( \frac{\pi \hbar}{e^2 R_\Box} \biggr) \,. \label{rate2D}
\end{equation}
Substituting our experimental value of $R_\Box$ into equation~(\ref{rate2D}), we obtain $A_{ee}^{{\rm
(2D,th)}}$ = 1.8$\times$$10^{10}$ K$^{-1}$ s$^{-1}$. Also, substituting our experimental values of
$E_F$ and $k_F$ into equation~(\ref{rate3D}), we obtain $A_{ee}^{{\rm (3D,th)}}$ = 2.7$\times$$10^9$
K$^{-3/2}$ s$^{-1}$. These values are in good agreement with the experimental values. Therefore, we
can clearly identify the $e$-$e$ scattering as the dominating dephasing process in the IZO2 NW. We
notice that our $A_{ee}^{\rm (2D)}$ value is on the same order of magnitude as that found in the ZnO
surface wells \cite{Goldenblum99}.

Our fitted $\tau_0$ value for the IZO2 NW is listed in table \ref{t3}. It should be noted that the
fitted $\tau_0$ value is only approximate, because we have not extensively measured the MR curves at
subkelvin $T$ values to unambiguously extract $\tau_0$ = $\tau_\varphi$($T$ $\rightarrow$ 0 K). We
also note that, at our lowest measurement temperature of 1 K, the electrons in the NW might have been
slightly overheated.

The electron dephasing processes in the IZO1 NW is somewhat more complicated and requires a more
detailed examination. Because both 1D-to-2D and 2D-to-3D dimensionality crossovers in the WL effect
are observed, we first write equation~(\ref{rate}) in the form $\tau_\varphi^{-1}$ = $\tau_0^{-1}$ +
$A_{ee}^{{\rm (2D)}}$$T$ + $A_{ee}^{{\rm (3D)}}$$T^{3/2}$ to extract the values of $A_{ee}^{{\rm
(2D)}}$ and $A_{ee}^{{\rm (3D)}}$ using data in the $T$ $\simeq$ 10--50 K interval. Then, we write
equation~(\ref{rate}) in the form $\tau_\varphi^{-1}$ = $\tau_0^{-1}$ + $A_{ee}^{{\rm (1D)}}$$T^{2/3}$
to extract the values of $\tau_0^{-1}$ and $A_{ee}^{{\rm (1D)}}$ using data in the $T$ $\simeq$ 2--10
K interval. Our fitted result in figure~\ref{fig7} is obtained with the following values:
$A_{ee}^{{\rm (1D)}}$ $\approx$ 1.6$\times$$10^{10}$ K$^{-2/3}$ s$^{-1}$, $A_{ee}^{{\rm (2D)}}$
$\approx$ 1.1$\times$$10^{10}$ K$^{-1}$ s$^{-1}$, and $A_{ee}^{{\rm (3D)}}$ $\approx$
2.3$\times$$10^{9}$ K$^{-3/2}$ s$^{-1}$. Substituting our experimental value of $R_\Box$ into
equation~(\ref{rate2D}), we obtain $A_{ee}^{{\rm (2D,th)}}$ = 2.4$\times$$10^{10}$ K$^{-1}$ s$^{-1}$.
Substituting our experimental values of $E_F$ and $k_F$ into equation~(\ref{rate3D}), we obtain
$A_{ee}^{{\rm (3D,th)}}$ = 9.8$\times$$10^8$ K$^{-3/2}$ s$^{-1}$. Our experimental values are within a
factor of $\approx$ 2 of the 2D and 3D theoretical values, and thus are satisfactory.

In 1D, the Nyquist $e$-$e$ scattering strength is theoretically predicted to be
\cite{Hsu-prb10,Altshuler85,Altshuler82}
\begin{equation}
A_{ee}^{{\rm (1D,th)}} = \Biggl( \frac{e^2 \sqrt{D} R k_B}{2 \sqrt{2} \hbar^2
L} \Biggr)^{2/3} \,. \label{rate1D}
\end{equation}
Substituting our measured NW resistance $R$, length $L$, and diffusion constant $D$ into
equation~(\ref{rate1D}), we obtain $A_{ee}^{{\rm (1D,th)}}$ $\approx$ 2.5$\times$$10^{10}$ K$^{-2/3}$
s$^{-1}$. This value is about 50\% higher than our experimental value, and hence our result is well
acceptable \cite{note10}.

\section{Conclusion}

We have measured the temperature dependence of resistance as well as the magnetic field dependence of
magnetoresistance in two indium-doped ZnO nanowires. The doped NWs reveal overall metallic transport
properties characteristic of disordered conductors. Our results lead to our proposition of a
core-shell-like structure in individual IZO NWs, with the outer shell of a thickness $\simeq$ 15--17
nm being responsible for the observed quantum-interference WL and EEI effects. As a consequence,
1D-to-2D and 2D-to-3D dimensionality crossovers in the WL effect are evident as the temperature
gradually increases from 1 to 70 K. A 2D-to-3D dimensionality crossover in the EEI effect has also
been observed. A crossover to the 1D EEI effect is not seen, because the thermal diffusion length $L_T
(T)$ is relatively short, as compared with the effective NW diameter $d$. These observations reveal
the complex and rich nature of the charge transport processes in group-III metal doped ZnO NWs. In
addition, we have explained the inelastic electron dephasing times. It should be emphasized that our
experimental observation of a core-shell-like structure in IZO NWs is in good accord with the current
theoretical understanding for impurity doping of semiconductor nanostructures. This result could have
significant bearing on the potential implementation of nanoelectronic devices.

\begin{acknowledgments}

This work was supported by the Taiwan National Science Council through Grant No NSC 100-2120-M-009-008 and by
the MOE ATU Program (JJL). Research by JGL was supported by NSF.\\

\noindent $^\ast$Email: jjlin@mail.nctu.edu.tw (Juhn-Jong Lin)

\end{acknowledgments}


\begin{thebibliography}{99}
\bibitem{x4} In the evaluation of $\rho$ in a given IZO NW, we have assumed that the electrical current passes
through the whole volume of a NW. If the electrical conduction is not through the bulk of a NW, the
above estimate of $\rho$ will be changed by an amount of $\lesssim$ 50\%, according to the geometries
of our core-shell-like structure. Since $k_F \propto n^{1/3}$, the values of the other electronic
parameters will thus be modified by smaller amounts of $\lesssim$ 15\%. Such modifications will not
affect any conclusions drawn in this work.
\bibitem{note5} Equation~(\ref{2D}) was originally formulated for a planar structure and with the $B$
field applied perpendicular to the film plane. In the present case, our conduction shell is roughly
cylindrical. A similar situation occurred in multiwalled carbon nanotubes, where equation~(\ref{2D})
had been successfully applied to describe the measured perpendicular MR in the 2D WL effect
\onlinecite{Langer-prl96,Tarkiainen-prb04}. In this work, by directly applying this equation also gave
satisfactory fitting results of $L_\varphi$ and $t$ for our IZO NWs. For example, we obtained $t$
$\simeq$ 17$\pm$2 nm for both NWs. Moreover, the fitted $L_\varphi$ values differed by less than
$\approx$ 10--15\% from those plotted in figures~\ref{fig4}(a) and \ref{fig4}(b). In a strict manner,
for a cylindrical film, the normal to the surface component of the $B$ field is not a constant but
depends on the position of the surface component under consideration. In this case, the normal
component of the applied $B$ field, $B_\perp$, on each surface component has to be taken into account.
In this work, we have approximated the cross section of the conduction shell in our NWs with a
dodecagon and explicitly taken the $B_\perp$ field into our least-squares fits. A similar approach has
recently been adopted in \onlinecite{Blomers-nanolett11} for their analysis of InAs NWs, where the
authors hypothesized a hexagonal cross section for their NWs.
\bibitem{note6} We note that a minute misalignment of the NW axis with respect to the $B_\|$ field would
cause an extra contribution from the perpendicular component of $B_\|$ to the measured parallel MR
data. In this experiment, we estimate that there could be a small misalignment of $<$ 5$^\circ$. Since
the WL MR curves of the IZO NWs are not dramatically anisotropic, any possible tiny erroneous
contribution from the perpendicular component of the $B_\|$ field should thus be insignificant in this
work.
\bibitem{note7} We notice that signatures of the AAS oscillations have recently been observed in radial
core/shell In$_2$O$_3$/InO$_x$ heterostructure NWs \onlinecite{Jung-nanolett08}. The authors found
that the electrical current flowed dominantly through the core/shell interface, but not through the
whole shell region. A very thin interface layer of sub-nanometer scale could sustain the AAS
oscillations.
\bibitem{note8} Apart from equations~(\ref{2Dee}) and (\ref{3Dee}), the EEI effect causes a
$-$$1/\sqrt{T}$ temperature dependence of the resistance rise in 1D, which is not seen in this study.
\end{thebibliography}
\end{document}